\documentclass[aps,prep]{revtex4}
\usepackage{amssymb}
\usepackage{amsmath}
\usepackage{bm}

\begin{document}

\title{Embedding General Relativity with varying cosmological constant term in five-dimensional Brans-Dicke theory of gravity in vacuum}
\author{$^{1}$ L. M. Reyes \thanks{%
E-mail address: luzreyes@ufpb.br}, and  $^{2,3}$ Jos\'e Edgar Madriz Aguilar \thanks{%
E-mail address: madriz@mdp.edu.ar} }
\affiliation{$^{1}$ Departamento de F\'isica, Universidade Federal da Para\'iba, C.P. 5008, CEP 58059-970, Jo\~{a}o Pessoa, PB, Brazil. \\ \\$^{2}$ Departamento de F\'isica, DCI, Campus Le\'on, Universidad de Guanajuato, C.P. 37150, Le\'on Guanajuato, M\'exico, and \\
$^3$ Departamento de F\'isica, Facultad de Ciencias Exactas y Naturales, Universidad Nacional de Mar del Plata, Funes 3350,
C.P. 7600, Mar del Plata, Argentina. \\
E-mail: luzreyes@ufpb.br; madriz@mdp.edu.ar}

\begin{abstract}
We investigate the possibility to recover a four-dimensional (4D) general theory of relativity, as embedded in a 5D spacetime where gravity is governed by a five-dimensional (5D) Brans-Dicke (BD) theory of gravity. Employing the Gauss-Codazzi-Ricci equations and some ideas of the induced matter theory, we obtain that when the 5D BD scalar field is only depending of the extra coordinate, it is possible to recover on every generic 4D hypersurface, the usual Einstein field equations  plus a cosmological constant term, with matter sources described by the energy momentum tensor of induced matter and where the cosmological constant, under certain conditions, can vary with time. Finally to illustrate the formalism, we give an example in which we induce a cosmological constant from warped product spaces.   
\end{abstract}

\maketitle

\vskip .5cm


\section{Introduction}
Currently theories of gravity in more than four dimensions have become a recourse of theoretical cosmologists to address among other cosmological issues resulting from the present evidences for dark energy. Among these theories we can count the brane world scenarios and the induced matter (IM) theory. The basic idea in the induced matter theory is that assuming a five-dimensional (5D) Ricci-flat spacetime, sources of matter in our four-dimensional (4D) spacetime can be geometrically induced or explained by using only one extended extra dimension \cite{wbook,rept}. Matematically this approach is supported by the Campbell-Magaard theorem which establishes that {\it any analytic $n$-dimensional Riemannian space can be locally embedded in a $(n+1)$-dimensional Ricci-flat space} \cite{CMT1,CMT2,CMT3,CMT4}. The ideas of the IM theory have been extended beyond this scenario to a more general theories of gravity as for instead the Brans-Dicke (BD) theory of gravity. The BD theory of gravity is an alternative to general relativity where in simple words the gravitational coupling is given as the inverse of a scalar field known as the BD scalar field \cite{BrD1,BrD2}. The BD theory of gravity and in general scalar tensor theories have been explored in the face of extra-dimensions (see for example \cite{ST1,ST2,ST3}). Recently it has been shown in \cite{BD5D} that when a 5D BD theory of gravity in vacuum is considered, it is possible to obtain a 4D modified version of the BD theory of gravity as embedded in the 5D spacetime. In this version classical matter configurations in 4D can be geometrically induced from 5D in an analog manner as it is usually done in the IM theory. \\    

The two first and more referenced brane world scenarios are the Arkani-Hamed-Dimopoulos-Dvali model \cite{AH1,AH2,AH3} and the Randall-Sundrum models I and II \cite{RS2,RS1}. The main motivation of the Randall-Sundrum model I was to propose a possible solution of the gauge hierarchy problem \cite{BWWG}. In general in the brane world framework our 4D spacetime is represented by a hypersurface called the brane embedded in a higher dimensional spacetime called the bulk \cite{PBrax}. The bulk not necessarily needs to be Ricci-flat, indeed in Randall-Sundrum brane world it is regarded an Anti-de-Sitter bulk \cite{RS2,RS1,BWWG}. One remarkable feature of theories in more than four dimensions is that its predictions on our 4D spacetime try to avoid conflicts with the general theory of relativity. Inspired in this fact and adopting some ideas of the IM theory, spacetimes with extra dimensions (or bulks in brane worlds) with non-Riemannian geometry in which a 4D spacetime with Riemannian geometry can be embedded, have been a topic recently explored (see for example \cite{DR}). The basic idea of this proposal is to allow the higher dimensional spacetime to have a non-Riemannian geometry as for instead a Weyl geometry but maintaining the Riemannian structure on every generic hypersurface defined by making the fifth coordinate a constant \cite{DR}. Leaving this idea to the cosmological scenario it is possible to induce on our 4D spacetime, gravitationally described by the general theory of relativity, a cosmological constant which is depending of the 5D Weyl scalar field \cite{WCC}.\\

In brane worlds particles described by the standard model of particle physics are confined on the brane while gravity and other exotic matter such as the superstring dilaton field can propagate in the bulk \cite{PBrax}. The fact that gravity propagates along the higher dimensional space (or the bulk in brane world models) is not exclusive of brane world scenarios, in general theories cataloged as non-compact Kaluza-Klein theories as for example the IM theory and its modifications share this property \cite{rept}. Moreover in models like the Randall-sundrum models this fact is preponderant to address the gauge hierarchy problem. Technically when gravity propagates in the higher dimensional spacetime it is possible to introduce a higher dimensional gravitational coupling, which many authors identify as the higher dimensional analogous of the Planck mass $M_p$. When a dimensional reduction mechanism is implemented then is possible to derive a relation between the higher dimensional gravitational coupling  and the 4D gravitational coupling sometimes represented by $M_{p}^{-2}$. In more simple words we can say then that in general the gravitational coupling in the higher dimensional spacetime is different than the one in the embedded 4D spacetime. However as with some exceptions there are not a consens on what kind of geometry or theory of gravity prevails on the higher dimensional spacetime  we can have an extended perspective of this idea. We can consider a higher dimensional spacetime where the gravitational coupling could varies along the higher dimensional spacetime but be a constant on the embedded 4D spacetime. A theory which in a natural manner involves a varying gravitational coupling is the Brans-Dicke theory of gravity. The idea of regarding a higher dimensional spacetime where gravity is governed by a Brans-Dicke theory has been already explored in theories as for example brane world models and String theories (see for example \cite{brBD51,brBD52,brBD53,brBD54}). In the ambit of extended versions of the IM theory it has been shown that starting from a 5D Brans-Dicke theory of gravity in vacuum it is possible to induce on the 4D embedded spacetime a modified Brans-Dicke theory of gravity \cite{BD5D}. Thus it seems suitable to regard a Brans-Dicke theory of gravity as a first attempt to describe a varying gravitational in the higher dimensional spacetime. However historically it is well known that in the face of observational evidences the general theory of relativity is more favored that the usual 4D Brans-Dicke theory. Thus it would be a suitable feature of this idea that besides of having a varying higher dimensional gravitational coupling which can be constant on the embedded 4D spacetime, if we could recover the general theory of relativity in 4D. Even more it could be better in order to possibly include dark energy scenarios to obtain on the 4D spacetime the general theory of relativity with an induced varying cosmological constant.  \\

In this letter employing some ideas of the induced matter theory we investigate the possibility of embedding the general theory of relativity with a varying cosmological constant term into a 5D Brans-Dicke  theory of gravity in vacuum. The paper is organized as follows. In section II we proceed to set up the Brans-Dicke theory in 5D and by means of the Gauss-Codazzi-Ricci we obtain the 4D induced field equations using a foliation of the fifth coordinate as a dimensional reduction mechanism. In subsection A of section I we obtain a relation between the gravitational coupling in 5D and the Planck mass in 4D. In section III we illustrate the formalism developed in the previous sections by inducing a 4D cosmological constant from 5D warped product spaces. Finally in section IV we give some final comments. \\

Our conventions are Latin indices run in the range $(0,1,...,4)$ with the exception of $i$ and $j$ that take values in the range $(1,2,3)$. Greek indices run from $(0,1,2,3)$. The metric signature we use is $(+,-,-,-,-)$. Finally we adopt units on which the speed of light $c=1$.

\section{Embedding 4D general relativity on 5D BD theory of gravity}

We start by considering a 5D spacetime $(M^{5},g_{ab})$ with a line element which can be splitted in the form (4+1) as  
\begin{equation}\label{a1}
dS^{2}=g_{ab} (\xi)d\xi^{a}d\xi^{b}=g_{\mu\nu}(x,\psi)dx^{\mu}dx^{\nu}+g_{\psi\psi}(x,\psi)d\psi^{2},
\end{equation}
with $g_{ab}$ being the 5D metric depending of the local coordinates $\lbrace \xi^{a}\rbrace=\lbrace x^{\alpha},\psi\rbrace$ where $\psi$ is the non-compact space-like fifth coordinate. Now we assume that gravity in 5D propagates according to a Brans-Dicke theory of gravity in vacuum i.e. there are no matter fields defined on the 5D spacetime. The 5D action that describes a BD theory of gravity in vacuum  reads \cite{BD5D}
\begin{equation}\label{a2}
^{(5)}{\cal S}=\int d^{5}\xi\sqrt{g}\left[\varphi\, ^{(5)}\!R-\frac{\omega}{\varphi}g^{ab}\varphi_{,a}\varphi_{,b}\right],
\end{equation}
where $g$ is the determinant of $g_{ab}$, the scalar field $\varphi (\xi)$ is the BD scalar field describing the gravitational coupling in 5D and $\omega$ is the BD constant parameter. Following \cite{BD5D} the field equations derived from this action can be written as 
\begin{eqnarray}\label{a3}
^{(5)}G_{ab}&=& \frac{\omega}{\varphi ^2}\left[\varphi _{,a}\varphi_{,b}-\frac{1}{2}g_{ab}\varphi^{,c}\varphi_{,c}\right]+\frac{1}{\varphi}\left[\varphi_{;a;b}-g_{ab}\varphi^{;c}\,_{;c}\right]\\
\label{a4}
^{(5)}\Box\varphi &=&0,
\end{eqnarray}
where the semicolon is denoting 5D covariant derivative and $^{(5)}\Box$ is denoting the 5D d'Alembertian operator. Inserting (\ref{a4}) in (\ref{a3}) we obtain
\begin{equation}\label{ab1}
^{(5)}G_{ab}= \frac{\omega}{\varphi ^2}\left[\varphi _{,a}\varphi_{,b}-\frac{1}{2}g_{ab}\varphi^{,c}\varphi_{,c}\right]+\frac{\varphi_{;a;b}}{\varphi}.
\end{equation}
Now let us to assume that the 5D spacetime can be foliated in such a way that for every generic hypersurface $\Sigma _{0}:\psi=\psi_0$ the induced metric is given by the 4D line element 
\begin{equation}\label{a5}
ds^{2}=h_{\alpha\beta}(x)dx^{\alpha}dx^{\beta},
\end{equation}
where the induced metric $h_{\alpha\beta}$ is given by $h_{\alpha\beta}=g_{\alpha\beta}(x,\psi_0)$. 
In analogy with the standard 4D BD theory of gravity \cite{Faraoni} the form of the action (\ref{a2}) or of the field equations (\ref{a3}) suggests that the BD scalar field $\varphi$ plays the role of the inverse of a gravitational coupling but the great difference is that in here this gravitational coupling give us the coupling of gravity on the higher dimensional spacetime. As it was shown in \cite{BD5D} if we maintain the field $\varphi$ depending on the whole coordinates when we go down to four dimensions under the approach of the induced matter theory it is obtained a modified BD theory of gravity on every hypersurface $\Sigma _{0}$. However if we would like to maintain a 4D gravitational coupling given by the Newtonian gravitational constant $G$ on the 4D spacetime, one election that we can make is to assume that the BD scalar field varies as a function only of the extra coordinate $\psi$. In this manner we are proposing that the gravitational coupling can vary along the fifth coordinate but it remains constant on every generic hypersurface $\Sigma _0$. Moreover as we will show this constant corresponds to the 4D gravitational coupling that can be fixed to be the Newtonian gravitational constant through the condition $\varphi (\psi_0)=1/(16\pi G)$. The interesting question here is to know what kind of theory will describe gravity on every hypersurface $\Sigma _0$ when $\varphi=\varphi (\psi)$. Now we will focus on answer this question and in order to do so we start by  expressing the field equations (\ref{ab1}) in the form (4+1) as
\begin{eqnarray}\label{a6}
^{(5)}G_{\alpha\beta}&=&\frac{1}{2}g^{\psi\psi}\left(\frac{\overset{\star}{\varphi}}{\varphi}\right)\left[\omega\left(\frac{\overset{\star}{\varphi}}{\varphi}\right)g_{\alpha\beta}-\overset{\star}{g}_{\alpha\beta}\right],\\
\label{a7}
^{(5)}G_{\alpha\psi}&=&-\frac{1}{2}\left(\frac{\overset{\star}{\varphi}}{\varphi}\right)g^{\psi\psi}g_{\psi\psi,\alpha},\\
\label{a8}
^{(5)}G_{\psi\psi}&=&\frac{\omega}{2}\left(\frac{\overset{\star}{\varphi}}{\varphi}\right)^{2}+\frac{\overset{\star\star}{\varphi}}{\varphi}-\frac{1}{2}\left(\frac{\overset{\star}{\varphi}}{\varphi}\right)g^{\psi\psi}\overset{\star}{g}_{\psi\psi},
\end{eqnarray}
where the star $(\star)$ is denoting derivative with respect to the fifth coordinate $\psi$. Under the assumption $\varphi=\varphi (\psi)$ the field equation (\ref{a4}) becomes
\begin{equation}\label{a9}
\frac{1}{\sqrt{g}}\frac{\partial}{\partial\psi}\left[\sqrt{g}\,g^{\psi\psi}\overset{\star}{\varphi}\right]=0.
\end{equation}
Now we can define in a way, from the field equations (\ref{ab1}), a 5D energy-momentum tensor given by
\begin{equation}\label{a10}
T^{(\varphi)}_{ab}=\frac{\omega}{\varphi ^2}\left[\varphi _{,a}\varphi_{,b}-\frac{1}{2}\,g_{ab}\varphi^{,c}\varphi_{,c}\right]+\frac{\varphi_{;a;b}}{\varphi}.
\end{equation}
Thus using some results of \cite{SeahraT} and the Gauss-Codazzi-Ricci equations (see for example \cite{GCR}) is easy to show that the dynamical field equations on $\Sigma _{0}$ have the form
\begin{equation}\label{a11}
^{(4)}G_{\alpha\beta}=\frac{1}{2\varphi(\psi_0)}T_{\alpha\beta}^{(\varphi)}-\frac{1}{2\varphi(\psi_0)}\left[h^{\mu\nu}T_{\mu\nu}^{(\varphi)}-\frac{2}{3}T^{(\varphi)}\right]h_{\alpha\beta}+\frac{1}{2\varphi(\psi_0)}\,T_{\alpha\beta}^{(IM)},
\end{equation} 
where $T^{(\varphi)}=g^{ab}T_{ab}^{(\varphi)}$ and $T^{(IM)}_{\alpha\beta}$ is the usual energy momentum tensor for induced matter which is given by the expression \cite{wbook,rept}
\begin{equation}\label{a12}
T^{(IM)}_{\alpha\beta}=g^{\psi\psi}(g_{\psi\psi ,\alpha})_{;\beta}-\frac{1}{2}(g^{\psi\psi})^{2}\left[g^{\psi\psi}\overset{\star}{g}_{\psi\psi}\overset{\star}{g}_{\alpha\beta}-\overset{\star\star}{g}_{\alpha\beta}+g^{\lambda\mu}\overset{\star}{g}_{\alpha\lambda}\overset{\star}{g}_{\beta\mu}-\frac{1}{2}\,g^{\mu\nu}\overset{\star}{g}_{\mu\nu}\overset{\star}{g}_{\alpha\beta}+\frac{1}{4}\,g_{\alpha\beta}\left(\overset{\star}{g}^{\mu\nu}\overset{\star}{g}_{\mu\nu}+(g^{\mu\nu}\overset{\star}{g}_{\mu\nu})^{2}\right)\right].
\end{equation}
Therefore when the BD scalar field $\varphi$ depends solely of the extra coordinate the induced field equations (\ref{a11}) yield
\begin{equation}\label{a13}
^{(4)}G_{\alpha\beta}=\Lambda(x)h_{\alpha\beta}+\frac{1}{2\varphi(\psi_0)}T_{\alpha\beta}^{(IM)},
\end{equation}
where $\Lambda(x)\equiv [\omega/(2\varphi(\psi_0)^{2})](g^{\psi\psi}\overset{\star}{\varphi}^{2})|_{\psi=\psi_0}$. These equations are very similar to the field equations of the usual 4D general relativity. In fact it can be easily seen by simple inspection that when $g_{\psi\psi}$ is a function of time or a constant then the function $\Lambda (x)$ could play the role of a time varying cosmological ``constant" or of a cosmological constant respectively. Matter configurations can be classically described  by the energy-momentum tensor of induced matter as it is usually done in the induced matter theory \cite{wbook,rept}. However there is one more difference between the equations (\ref{a13}) and the Einstein's field equations of general relativity and it is that in Einstein's general relativity the gravitational coupling is given by the Newton's constant $G_{N}$. This difference can be eliminated by requiring the condition $\varphi(\psi_0)=1/(16\pi G_{N})$. Therefore we can say that when $g_{\psi\psi}$ is a function at most of the time coordinate $t$ and of the extra coordinate $\psi$, and when we have $\varphi(\psi_0)=1/(16\pi G_{N})$ we can recover on every hypersurface $\Sigma _0$ of the foliation the Einstein's field equations of general relativity with a cosmological ``constant" term that could be in principle time depending and in this sense we have shown that when the Brans-Dicke scalar field $\varphi$ is only depending of the fifth coordinate, it is possible to see 4D general relativity as embedded in a 5D Brans-Dicke theory of gravity in vacuum.

\subsection{On the Planck mass and the higher dimensional gravitational coupling}

The gauge hierarchy problem basically consists in explaining why is so large the mass gap between the Planck scale ($M_{Planck}=1.22\cdot10^{19}$ Gev) and the electroweak scale ($M_{EW}\sim 10^{3}$ Gev). A first step in some higher dimensional theories, as for instead the Randall and Sundrum model I \cite{RS1,RS2}, to address this problem is to derive a relation between the Planck mass and the mass scale of the higher dimensional theory via a dimensional reduction \cite{BWWG}. In more simple words they establish a relation between the higher dimensional gravitational coupling and the 4D one. Here we will derive that relation in a similar manner as it is done in \cite{BWWG}.\\ 

We start by constructing the effective 4D action on the hypersurface $\Sigma _{0}$ which reads
\begin{equation}\label{a14}
^{(4)}{\cal S}=\int d^{4}x\int d\psi\delta (\psi-\psi_0) \frac{\sqrt{g}}{\sqrt{-g_{\psi\psi}}}\left[\varphi\,^{(5)}R - \frac{\omega}{\varphi}g^{ab}\varphi_{,a}\varphi_{,b}\right],
\end{equation}
where since we are regarding the fifth $\psi$ dimension as non compact the integration limits for the integration over $\psi$ go from $-\infty$ to $+\infty$. The factor $1/\sqrt{g_{\psi\psi}}$ has been introduced in order to have the real differential element of volume in 4D. Now using the fact that geometrically $^{(5)}R=\,^{(4)}R -(K^{\mu\nu}K_{\mu\nu}-K^{2})$ with $K_{\mu\nu}$ being the extrinsic curvature tensor and $K=h^{\alpha\beta}K_{\alpha\beta}$, and performing the integration over $\psi$, the action (\ref{a14}) can be written as 
\begin{equation}\label{a15}
^{(4)}{\cal S}=\int d^{4}x\, \sqrt{-h}\left[\varphi(\psi_0)\,^{(4)}R+{\cal L}_{IM}-2\varphi(\psi_0)\Lambda(x)\right],
\end{equation}
where ${\cal L}_{IM}=\varphi(\psi_0)(K^{2}-K^{\mu\nu}K_{\mu\nu})$ is the geometrical lagrangian density which generates the energy-momentum tensor of  induced matter $T_{\alpha\beta}^{(IM)}$ and as it was defined previously $\Lambda(x)=[\omega/(2\varphi(\psi_0)^{2})](g^{\psi\psi}\overset{\star}{\varphi}\,^{2})|_{\psi=\psi_0}$. The resulting action (\ref{a15}) prescribes the gravitational dynamics on the hypersurface $\Sigma _0$. Taking into account that our interest is to recover the general theory of relativity on $\Sigma _0$, we compare the action (\ref{a15}) with the Einstein-Hilbert action of the general theory of relativity with a cosmological constant term. Hence according to the first term in (\ref{a15}) we obtain the relation 
\begin{equation}\label{a16}
M_{Pl}^{2}=16\pi\varphi (\psi _0).
\end{equation}
This relation means that in order to recover general relativity with a varying cosmological constant term on $\Sigma_0$, the gravitational coupling in 5D evaluated on $\Sigma _0$ must be given in terms of the Planck mass. Moreover note that even when the $\Lambda (x)$ obtained from (\ref{a15}) was derived independently of the one derived in (\ref{a13}), they are the same. This means that field equations (\ref{a13}) can be completely derived from the action (\ref{a15}).

\section{The induced 4D cosmological constant from warped product spaces}

In order to illustrate how the geometrical mechanism works to induce a cosmological constant on the 4D spacetime $\Sigma _0$ we consider the 5D line element 
\begin{equation}\label{b1}
dS^{2}=e^{2A(\psi)}[dt^{2}-a^{2}(t)\delta_{ij}dx^{i}dx^{j}]-d\psi^{2},
\end{equation}
where $t$ is the cosmic time, $A(\psi)$ is a warping factor, $a(t)$ is the cosmological scale factor and $\delta_{ij}$ is the Kronecker delta. On this geometrical background the field equations (\ref{a6}) to (\ref{a8}) are given by
\begin{eqnarray}
\label{b2}
3H^{2}-3(\overset{\star}{A}^{2}+\overset{\star\star}{A})e^{2A}&=& \frac{1}{2}\frac{\overset{\star}{\varphi}}{\varphi}\left(\omega\frac{\overset{\star}{\varphi}}{\varphi}-2\overset{\star}{A}\right)e^{2A},\\
\label{b3}
2\frac{\ddot{a}}{a}+H^{2}-3(\overset{\star}{A}^{2}+\overset{\star\star}{A})e^{2A}&=&\frac{1}{2}\frac{\overset{\star}{\varphi}}{\varphi}\left(\omega\frac{\overset{\star}{\varphi}}{\varphi}-2\overset{\star}{A}\right)e^{2A},\\
\label{b4}
-3\frac{\ddot{a}}{a}-3H^{2}+6\overset{\star}{A}^{2}e^{2A}&=&\frac{1}{2}\left[\omega\left(\frac{\overset{\star}{\varphi}}{\varphi}\right)^{2}+\frac{\overset{\star\star}{\varphi}}{\varphi}\right]e^{2A}.
\end{eqnarray}
The equations (\ref{b2}), (\ref{b3}), (\ref{b4}) correspond respectively to the components $^{(5)}G_{tt}$, $^{(5)}G_{ii}$ and $^{(5)}G_{\psi\psi}$. Thus the combination $^{(5)}G_{tt}+3\,^{(5)}G_{ii}+2\,^{(5)}G_{\psi\psi}$ yields 
\begin{equation}\label{b5}
2\varphi\overset{\star\star}{\varphi}-4\overset{\star}{A}\varphi\overset{\star}{\varphi}+3\omega\overset{\star}{\varphi}^{2}+12\overset{\star\star}{A}\varphi ^{2}=0.
\end{equation}
The equation (\ref{a9}) under the geometrical background (\ref{b1}) gives for solution
\begin{equation}\label{b6}
\varphi(\psi)=k_{\psi}\int_{\psi_0}^{\psi}\exp{[-4A(\psi\prime)]}\,d\psi\prime +\varphi_{0},
\end{equation}
where $\varphi _{0}=\varphi(\psi_0)$ and $k_{\psi}$ is a separation constant. Both equations (\ref{b5}) and (\ref{b6}) determine the Brans-Dicke scalar field $\varphi$ as a function of the warping factor $A(\psi)$. Hence for a warping of the form 
\begin{equation}\label{b7}
A(\psi)=n\,\ln\left[\frac{\psi_0}{\psi+\psi_0}\right],
\end{equation}
being $n>0$, the expression (\ref{b6}) leads to a Brans-Dicke scalar field
\begin{equation}\label{b8}
\varphi(\psi)=\frac{1}{4n+1}\frac{k_\psi}{\psi_{0}^{4n}}\,(\psi+\psi_0)^{1+4n}\,.
\end{equation}
Inserting the equation (\ref{b8}) in (\ref{b5}) it gives that in order to (\ref{b5}) to be satisfied for the field (\ref{b8}) necessarily 
\begin{equation}\label{b9}
\omega=-\frac{8n(2n+1)}{(4n+1)^{2}}.
\end{equation}  
On the other hand in order to the condition $\varphi(\psi _0)=1/(16\pi G_N)$ holds, the separation constant in (\ref{b8}) must be given by $k_\psi=[(4n+1)/(16\pi G_N)][1/(\psi_{0}2^{4n+1})]$ and thus the expression (\ref{b8}) finally becomes
\begin{equation}\label{b10}
\varphi(\psi)=\frac{1}{16\pi G_{N}}\left(\frac{\psi +\psi_0}{2\psi_{0}}\right)^{1+4n}.
\end{equation}
The effective 5D gravitational coupling is given as in the usual Brans-Dicke theory of gravity by $^{(5)}G_{eff}=1/\varphi$ and hence according to  (\ref{b10}) the 5D effective gravitational coupling reads
\begin{equation}\label{b11}
G_{eff}(\psi)\equiv\frac{1}{\varphi(\psi)}=16\pi G_{N}\left(\frac{2\psi_0}{\psi+\psi_0}\right)^{1+4n}\,.
\end{equation}
The induced metric (\ref{a5}) on the generic 4D hypersurface $\Sigma _0$ derived from (\ref{b1}) can be written as 
\begin{equation}\label{be1}
ds^{2}=dT^{2}-a^{2}(T)\delta_{ij}dX^{i}dX^{j},
\end{equation}
where we have made the space-time coordinate rescaling $dT=e^{A_0}dt$, $dX^{i}=e^{A_0}dx^{i}$. Clearly this metric corresponds to a FRW metric with a scale factor $a(t)$. According to (\ref{a13}) and employing the equations (\ref{b9}) and (\ref{b10}) the induced 4D cosmological constant will be in this case
\begin{equation}\label{b12}
\Lambda_{0}=\frac{n(1+2n)}{\psi_{0}^{2}}.
\end{equation}
The induced matter on 4D is described classically by the energy momentum tensor given by (\ref{a12}), however as it is usually done in cosmological applications on the framework of the induced matter approach \cite{wbook,rept}, we can regard matter in 4D as a perfect fluid through an energy momentum tensor of the form $T_{\mu\nu}=(\rho_{(IM)}+p_{(IM)})u_{\mu}u_{\nu}-p_{(IM)}h_{\mu\nu}$ with the 4-velocity of the observers given by $u^{\mu}=\delta^{\mu}_{0}$ (comoving observers located on $\Sigma _0$) and satisfying $u^{\mu}u_{\mu}=1$. Thus according to (\ref{a13}) the field equations on $\Sigma_0$ are
\begin{eqnarray}\label{e4d}
3H^{2}&=&8\pi G\rho _{eff},\\
\label{ee4d}
2\frac{\ddot{a}}{a}+H^{2}&=&-8\pi G p_{eff},
\end{eqnarray}
where we have defined the effective energy density and pressure by $\rho _{eff}=\rho_{(IM)}+\Lambda _0$ and $p_{eff}=p_{(IM)}-\Lambda _0$.
In terms of the warping factor $A(\psi)$ the energy density $\rho _{(IM)}$ and the pressure $p_{(IM)}$ of induced matter  are determined by
\begin{equation}\label{b13}
\rho_{(IM)}=\overset{\star\star}{A}-2\overset{\star}{A}^{2},\qquad p_{(IM)}=-(\overset{\star\star}{A}-2\overset{\star}{A}^{2}) ,
\end{equation}
where both quantities must be evaluated at $\psi=\psi_0$. Using the equation (\ref{b7}) the previous $\rho_{(IM)}$ and $p_{(IM)}$ become
\begin{equation}\label{b14}
\rho_{(IM)}=\frac{n(1-2n)}{4\psi_0^2},\qquad p_{(IM)}=-\frac{n(1-2n)}{4\psi_0^2}.
\end{equation}
It can be easily seen from the first of these equations that in order to have a positive energy density $\rho_{(IM)}$ necessarily $0<n<1/2$. Taking into account this restriction and in order to warranty a positive value of $\varphi$ for every value of $\psi$ in the equation (\ref{b10}), $n=1/4$ seems to be a convenient election of $n$. In this case the Brans-Dicke scalar field given by the equation (\ref{b10}) and consequently the effective 5D gravitational coupling determined by (\ref{b11}) become respectively
\begin{eqnarray}\label{b15}
\varphi(\psi)&=&\frac{1}{16\pi G_{N}}\left(\frac{\psi +\psi_0}{2\psi_{0}}\right)^{2},\\
\label{b16}
G_{eff}(\psi)&=&16\pi G_{N}\left(\frac{2\psi_0}{\psi+\psi_0}\right)^{2}.
\end{eqnarray}
The effective equation of state parameter $\omega _{eff}$ in this case gives
\begin{equation}\label{b17}
\omega _{eff}\equiv\frac{p_{eff}}{\rho_{eff}}=\frac{p_{(IM)}-\Lambda_{0}}{\rho_{(IM)}+\Lambda_{0}}=-1,
\end{equation}
which corresponds to an equation of state for a purely cosmological constant. Finally the 4D induced cosmological constant according to (\ref{b12}) for $n=1/4$ is given by
\begin{equation}\label{b18}
\Lambda_{0}=\frac{3}{8\psi_{0}^{2}}.
\end{equation}
The present value of the cosmological constant according to cosmological estimations is approximately $\Lambda _{0}=10^{-54}\,m^{-2}$ or $\Lambda=10^{-47}\,GeV^{4}$ \cite{Pad}, thus when we impose this restriction to (\ref{b18}) it can be easily shown that $\psi _{0}=6.12\cdot 10^{26}\, m$. This value is approximately of the order of the present Hubble length $\lambda _{H_0}\simeq 1.27\cdot 10^{26}\,m$ and it could explain why for  observers located on $\Sigma_0$ the extended extra coordinate remains unobserved at least directly.

\section{Final Comments}

In this letter we have shown how 4D general relativity with a cosmological constant term can be embedded in a 5D space-time where gravity is governed by a Brans-Dicke theory of gravity. The main condition in order to the embedding be possible is that the five-dimensional Brans-Dicke scalar field $\varphi$ solely depends of the extra coordinate $\psi$. In general a natural feature of a Brans-Dicke theory of gravity is that the gravitational coupling is described by the Brans-Dicke scalar field which makes that coupling variable in every point of the space-time. In this letter we take advantage of this property to obtain a five-dimensional gravitational coupling that could varies along the extra dimension but remains constant on every generic hypersurface defined by the foliation $\psi=\psi_0$. Employing the Gauss-Codazzi-Ricci equations and making use of the former property, we showed that it is possible to obtain on every hypersurface $\Sigma_{0}:\psi=\psi_0$ the field equations of the usual 4D general theory of relativity with varying cosmological constant term, where both the cosmological constant term and the matter can be induced from the 5D space-time via a geometrical mechanism. Moreover the 4D gravitational coupling which in general relativity is given by the Newton's constant $G_N$ is also induced from the higher dimensional gravity, in particular $16\pi G_{N}=1/\varphi(\psi _0)$. Depending of the 5D metric the induced cosmological term gives the option of inducing a cosmological constant $\Lambda _0$ or a time varying cosmological ``constant'' $\Lambda (t)$. The first case is obtained when 5D metric component $g_{\psi\psi}$ is a constant and the second one when $g_{\psi\psi}$ is time dependent. \\

By constructing the effective 4D action (\ref{a14}) from the 5D Brans-Dicke action (\ref{a2}) we derived a relation between the 5D gravitational coupling evaluated on $\psi=\psi_0$ and the Planck mass, specifically we obtain that $M_{p}^{2}=16\pi\varphi_{0}$. This result combined with the induced field equations (\ref{a13}), is interpreted as gravity on $\Sigma _0$ can be described by the general theory of relativity with a cosmological constant term, where the 4D coupling constant given by the Newton's constant $G_N$, has its origin on the 5D Brans-Dicke scalar field.\\

Finally in order just to illustrate how the geometrical mechanism of inducing the cosmological constant term works, we have derived a cosmological constant for the class of geometries known as warped product geometries. We derived a 4D cosmological scenario where matter is geometrically induced in the same way as it is usually induced on the induced matter approach. The equation of state parameter for induced matter is $\omega_{IM}=-1$ leading when the cosmological constant is included to an effective equation of state parameter $\omega _{eff}=-1$. The fact that $\omega _{IM}=-1$ is an inherent property of the warped product geometries and hence in a case we would like to obtain a more general or realistic equation of state for induced matter, a 5D metric not belonging to this family of metrics must be considered. In this example in order to have a cosmological constant compatible with the cosmological observations ($\Lambda=10^{-54}\,m^{-2}$ or $\Lambda=10^{-47}\,GeV^{4}$) the fifth coordinate evaluated on $\Sigma _0$ must be of the order of the present Hubble horizon $\lambda _{H_0}\simeq 1.27\cdot 10^{26}\,m$, indeed both length scales satisfy $\lambda_{H_0}/\psi_0\simeq 0.2$. This fact could explain in this example why for hypothetical observers located on $\Sigma _0$  the extended fifth dimension remains unobserved at least directly.

\section*{Acknowledgements}

\noindent L. M. R. acknowledges CLAF-CNPq Brazil and J.E.M.A acknowledges CONACYT M\'exico  for financial support.

\end{document}